# Real-time image processing with a 2D semiconductor neural network vision sensor

Lukas Mennel[‡], Joanna Symonowicz, Stefan Wachter, Dmitry K. Polyushkin, Aday J. Molina-Mendoza, and Thomas Mueller[‡]

*Vienna University of Technology, Institute of Photonics, Gußhausstraße 27-29, 1040 Vienna, Austria*

[‡] Corresponding authors: lukas.mennel@tuwien.ac.at, thomas.mueller@tuwien.ac.at

**In recent years, machine vision has taken huge leaps and is now becoming an integral part of various intelligent systems, including autonomous vehicles, robotics, and many others. Usually, visual information is captured by a frame-based camera, converted into a digital format, and processed afterwards using a machine learning algorithm such as an artificial neural network (ANN) [1]. A large amount of (mostly redundant) data being passed through the entire signal chain, however, results in low frame rates and large power consumption. Various visual data preprocessing techniques have thus been developed [2,3,4,5,6,7] that allow to increase the efficiency of the subsequent signal processing in an ANN. Here, we demonstrate that an image sensor itself can constitute an ANN that is able to simultaneously sense and process optical images without latency. Our device is based on a reconfigurable two-dimensional (2D) semiconductor [8,9] photodiode [10,11,12] array, with the synaptic weights of the network being stored in a continuously tunable photoresponsivity matrix. We demonstrate both supervised and unsupervised learning and successfully train the sensor to classify and encode images, that are optically projected onto the chip, with a throughput of 20 million bins per second.**

ANNs have achieved huge success as a machine learning algorithm in a wide variety of fields [1]. The computational resources required to perform machine learning tasks are very demanding. Accordingly, dedicated hardware solutions, that provide better performance and energy efficiency than conventional computer architectures, have become a major focus of research. However, while much progress has been made in efficient neuromorphic processing of electrical [13,14,15,16] or optical [17,18,19,20] signals, the conversion of optical images into the electrical domain remains a bottleneck, particularly in time-critical applications. Imaging systems that mimic neuro-biological architectures may allow to overcome these disadvantages. Much work has therefore been devoted to develop systems that emulate certain functions of the human eye [21], including hemispherically shaped image sensors [22,23] and preprocessing of visual data [2,3,4,5,6,7], for example, for image contrast enhancement, noise reduction, or event-driven data acquisition.

Here, we present a photodiode array which itself represents an ANN that simultaneously senses and processes images projected onto the chip. The sensor performs a real-time multiplication of the projected image with a photoresponsivity matrix. Training of the network requires to set the photoresponsivity value of each pixel individually. Conventional photodiodes based, for example, on silicon exhibit a fixed responsivity, that is given by the inner structure (chemical doping profile) of the device, and are thus not suitable for the proposed application. We have therefore chosen $WSe_2$ – a 2D semiconductor – as photoactive material. 2D semiconductors not only show strong light-matter interaction and excellent optoelectronic properties [8,9], but also offer the possibility of external tunability of the potential profile in a device – and hence its photosensitivity – by electrostatic doping using multi-gate electrodes [10,11,12]. At the same time, 2D material technology has by now achieved a sufficiently high level of maturity to be employed in complex systems [24] and provides ease of integration with silicon readout/control electronics [25].

Fig. 1a schematically illustrates the basic layout of the image sensor. It consists of $N$ photoactive pixels, arranged in a two-dimensional array, with each pixel being divided into $M$ subpixels. Each subpixel is composed of a photodiode, which is operated under short-circuit conditions and, under optical illumination, delivers a photocurrent $I_{mn} = R_{mn}E_n A = R_{mn}P_n$, where $R_{mn}$ is its photoresponsivity, $E_n$ and $P_n$ denote the local irradiance and optical power at the $n^{th}$ pixel, respectively, and $A$ is the detector area. $n = 1, 2, ... N$ and $m = 1, 2, ... M$ denote the pixel and subpixel indices. An integrated neural network and imaging array can now be formed by interconnecting the subpixels. Summing up all photocurrents produced by the $m^{th}$ detector element of each pixel,

$$I_m = \sum_{n=1}^{N} I_{mn} = \sum_{n=1}^{N} R_{mn}P_n, \qquad (1)$$

performs a matrix-vector product operation $\boldsymbol{I} = \boldsymbol{R} \cdot \boldsymbol{P}$, with $\boldsymbol{R} = (R_{mn})$ being the photoresponsivity matrix, $\boldsymbol{P} = (P_1, P_2 ... P_N)^T$ a vector that represents the optical image projected onto the chip, and with currents $\boldsymbol{I} = (I_1, I_2 ... I_M)^T$ that form the output vector.

Provided that $R_{mn}$ of each detector element can be set to a specific positive *or negative* value, various types of ANNs for image processing can be implemented (see Figs. 1c, d), with the synaptic weights being encoded in the photoresponsivity matrix. The expression 'negative photoresponsivity' is to be understood in this context as referring to the sign of the photocurrent.

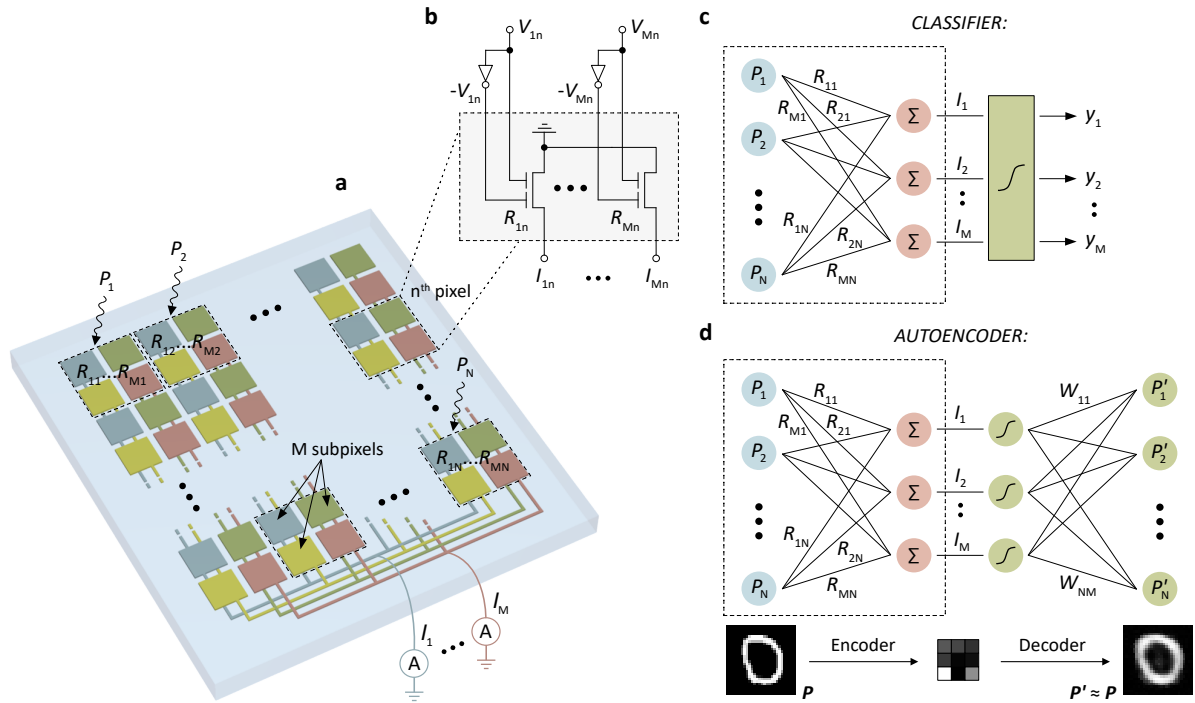

**Figure 1 | Imaging ANN photodiode array. a**, Illustration of the ANN photodiode array. All subpixels with same color are connected in parallel in order to generate $M$ output currents. **b**, Circuit diagram of a single pixel in the photodiode array. **c/d**, Schematic of the classifier/autoencoder. Below the illustration of the autoencoder an example of an encoding/decoding of a 28×28 pixel letter from the MNIST handwritten digit database is shown. The original image is encoded to 9 code layer neurons and then decoded back into an image.

In this work, we implemented two types of ANNs: a classifier and an autoencoder. Fig. 1c shows the schematic of a classifier. Here, the array is operated as a single-layer perceptron, together with nonlinear activation functions that are implemented off-chip. This type of ANN represents a supervised learning algorithm that is capable of classifying images $P$ into different categories $y$. An autoencoder (Fig. 1d) is an ANN that can learn, in an unsupervised training process, an efficient representation (encoding) for a set of images $P$. Along with the encoder, a decoder is trained so that it attempts to reproduce at its output the original image, $P' \approx P$, from the compressed data. The encoder is here formed by the photodiode array itself and the decoder again by external electronics.

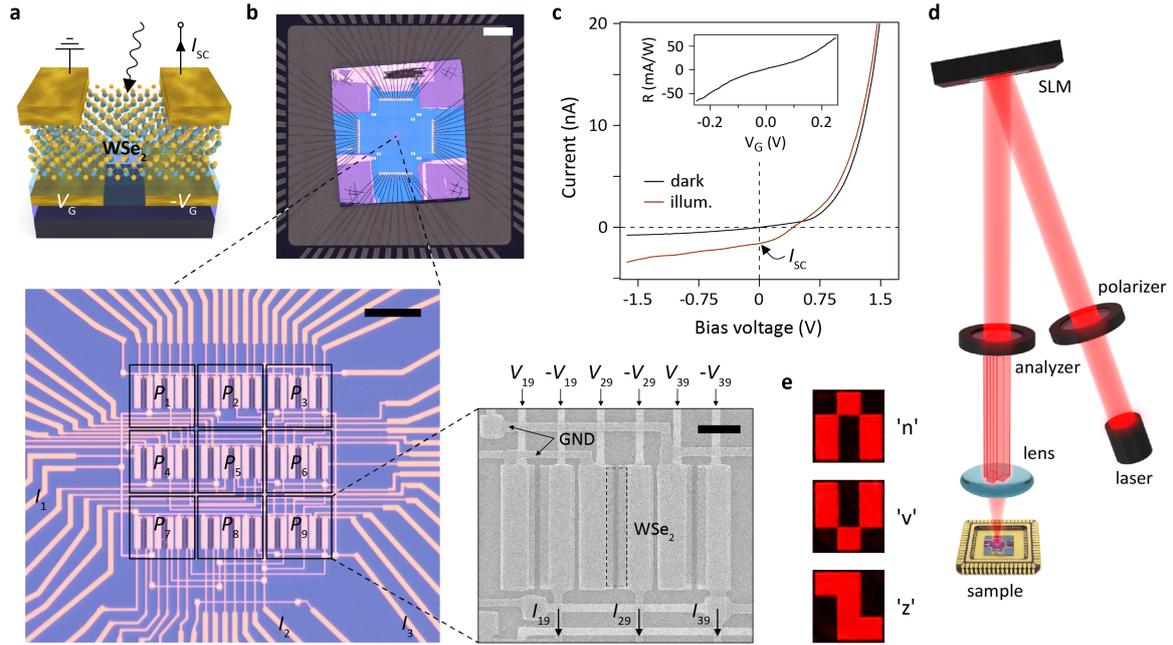

**Figure 2 | Implementation of the ANN photodiode array. a**, Schematic of a single WSe$_2$ photodiode. **b**, Macroscopic image of the bonded chip on the chip carrier. Scale bar, 2 mm. First magnification: Microscope image of the photodiode array, consisting of 3 × 3 pixels. Scale bar, 15 μm. Second magnification: Scanning electron microscopy (SEM) image of one of the pixels. Each pixel consists of three WSe$_2$ photodiodes/subpixels with responsivities set by the gate voltages. Scale bar, 3 μm. **c**, Current-voltage characteristic of one of the photodetectors in dark (blue line) and under optical illumination (red line). The inset shows the gate voltage tunability of the photoresponsivity. **d**, Schematic illustration of the optical setup. Laser light is linearly polarized by a wire-grid polarizer and reflected by a spatial light modulator (SLM). The reflected light is then filtered by an analyzer (intensity modulation) and the resulting image is projected onto the photodiode array. **e**, Microscope images of the 3×3 pixel letters used for training/operation of the network.

Having presented the operational concept of our network, we now come to an actual device implementation. We used a few-layer WSe$_2$ crystal with a thickness of ~4 nm to form lateral p–n junction photodiodes, using split-gate electrodes (with a ~300 nm wide gap) that couple to two different regions of the 2D semiconductor channel (Fig. 2a) [10,11,12]. WSe$_2$ was chosen because of its ambipolar conduction behavior and excellent optoelectronic properties. Biasing one gate electrode with $V_G$ and the other with $-V_G$ allows for adjustable (trainable) responsivities between -60 and +60 mA/W, as shown in Fig. 2c. This technology was then used to fabricate the photodiode array in Fig. 2b, consisting of a total of 27 detectors, with good uniformity and linearity (see Supplementary Fig. S1). The devices were arranged to form a 3×3 imaging array ($N = 9$), with a pixel size of ~17×17 μm$^2$ and with three detectors per pixel ($M = 3$). The short-circuit photocurrents $I_{SC}$, produced by the individual devices under optical illumination, were summed up according to Kirchhoff's law by hard-wiring devices in parallel, as depicted in Fig. 1b. The sample fabrication is explained in the Methods section and the schematic of the entire circuit is provided in Supplementary Fig. S2. Each device was supplied with a gate voltage pair $V_G/-V_G$ to individually set its responsivity. For training and testing of the chip, optical images were projected using the setup shown in Fig. 2d (for details, see Methods section). Unless otherwise stated, all measurements were performed with light at 650 nm wavelength and with a maximum irradiance of ~0.1 W/cm². Despite its small size, such a network is sufficient for the proof-of-principle demonstration of several machine learning algorithms. In particular, we performed classification, encoding, and denoising of the stylized letters 'n', 'v' and 'z' depicted Fig. 2e. Scaling the network to larger dimensions is conceptually straight-forward and remains a mainly technological task.

To test the functionality of the photodiode array, we first operated it as a classifier (Fig. 1c) to recognize the letters 'n', 'v' and 'z'. During each training epoch we optically projected a set of $S = 20$ randomly chosen letters. Gaussian noise (with standard deviation of $\sigma = 0.2$, 0.3 and 0.4; Fig. 3c) was added to augment the input data [26]. In this supervised learning example, we chose one-hot encoding, in which each of the three letters activates a single output neuron. We therefore used a softmax activation function for the $M$ photocurrents $\phi_m(I) = e^{I_m \xi}/\sum_{k=1}^{M} e^{I_k \xi}$, where $\xi = 10^{10}$ A$^{-1}$ is a scaling factor that ensures that the full value range of the activation function is accessible during training. As a loss function we used cross-entropy $\mathcal{L} = -\frac{1}{M}\sum_{m=1}^{M} y_m \log[\phi_m(I)]$, where $y_m$ is the label and $M = 3$ is the number of classes. The activations of the output neurons represent the probabilities for each of the letters. The responsivities were updated after each epoch by backpropagation [27] of the gradient of the loss function,

$$R_{mn} \rightarrow R_{mn} - \frac{\eta}{S}\sum_{P} \nabla_{R_{mn}}\mathcal{L}, \quad (2)$$

with learning rate $\eta = 0.1$.

In Figs. 3a and b accuracy and loss are plotted over the epochs. The loss is decreasing quickly for all noise levels and reaches a minimum after 15, 20, and 30 epochs for $\sigma = 0.2$, $\sigma = 0.3$, and $\sigma = 0.4$, respectively. The accuracy reaches 100% for all noise levels, with faster convergence for less noise. In Fig. 3d we show the currents for each of the three letters for $\sigma = 0.2$ (see Supplementary Fig. S3 for the other cases). The currents become well-separated after ~10 epochs, with the highest current corresponding to the label of the projected letter. The inset in Fig. 3b shows histograms for the (randomly chosen) initial and final responsivity values for $\sigma = 0.2$. The converged responsivity values for the other noise levels can be found in Supplementary Fig. S3.

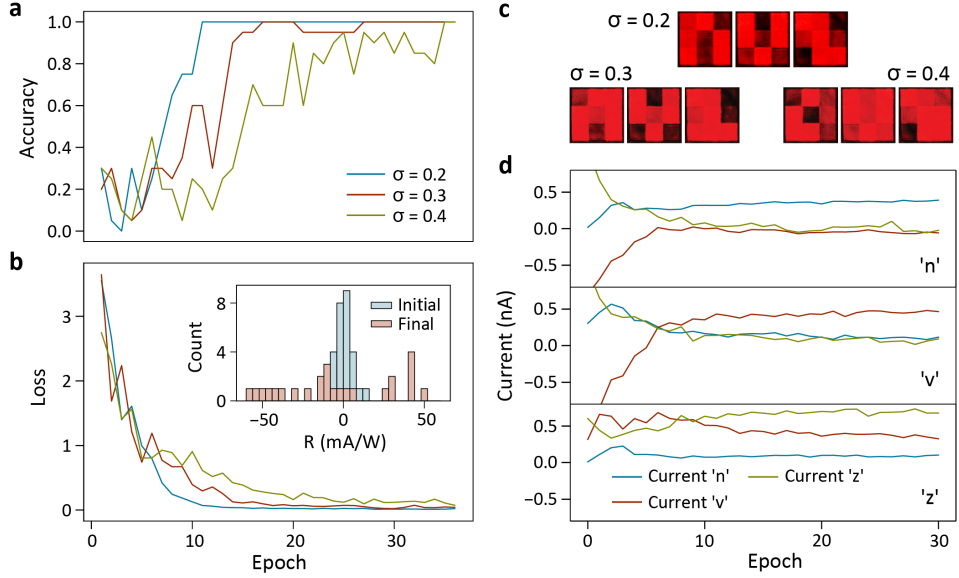

**Figure 3 | Device operation as a classifier. a**, Accuracy over epoch during training of the classifier for varying artificial noise levels. An image is accurately predicted when the correct neuron has the largest activation. **b**, Value of the loss function plotted for different noise levels over the epochs in the training. The inset shows the initial and final responsivity distributions for $\sigma = 0.2$. **c**, Microscope images of the projected letters with varying added random noise levels. **d**, Measured average currents for each epoch for a specific projected letter, for training with a noise level of $\sigma = 0.2$.

Next, we demonstrate encoding of image patterns by our device operating as an autoencoder (Fig. 1d). We chose logistic (sigmoid) activation functions for the code neurons $\phi_m(I_m) = (1 + e^{-I_m \xi})^{-1}$, again with $\xi = 10^{10}$ A$^{-1}$ as a scaling factor, as well as for the output neurons $P'_n = \phi_n(z_n) = (1 + e^{-z_n})^{-1}$, where $z_n = \sum_{n=1}^{N} W_{nm} \phi_m(I_m)$. $W_{nm}$ denotes the weight matrix of the decoder. We used a mean-squared loss function, $\mathcal{L} = \frac{1}{2} \|\boldsymbol{P} - \boldsymbol{P}'\|^2$, which depends on the difference between the original and reconstructed images. The responsivities were again trained by backpropagation of the loss according to Equation (2), with a noise level of $\sigma = 0.15$. Along with the encoder responsivities, the weights of the decoder $W_{nm}$ were trained. As shown in Fig. 4a, the loss steeply decreases within the first ~10 training epochs and then slowly converges to a final value after ~30 epochs. The initial and final responsivities/weights of the encoder/decoder are shown in Fig. 4b and the coded representations for each letter are visualized in Fig. 4c. Each projected letter delivers a unique signal pattern at the output. A projected 'n' delivers negative currents to code layer neuron 1 and 2 and a positive current to code layer neuron 3. After the sigmoid function, this causes only code layer neuron 3 to deliver a significant signal. The letters 'v' and 'z' activate two code layer neurons: 'v', code layer neuron 0 and 2; 'z' code layer neuron 1 and 2. The decoder transforms the coded signal back into an output that correctly represents the input. To test the fault tolerance of the autoencoder, we projected twice as noisy ($\sigma = 0.3$) images. Not only does the autoencoder interpret the inputs correctly, but the reconstructions are significantly less noisy (Fig. 4d).

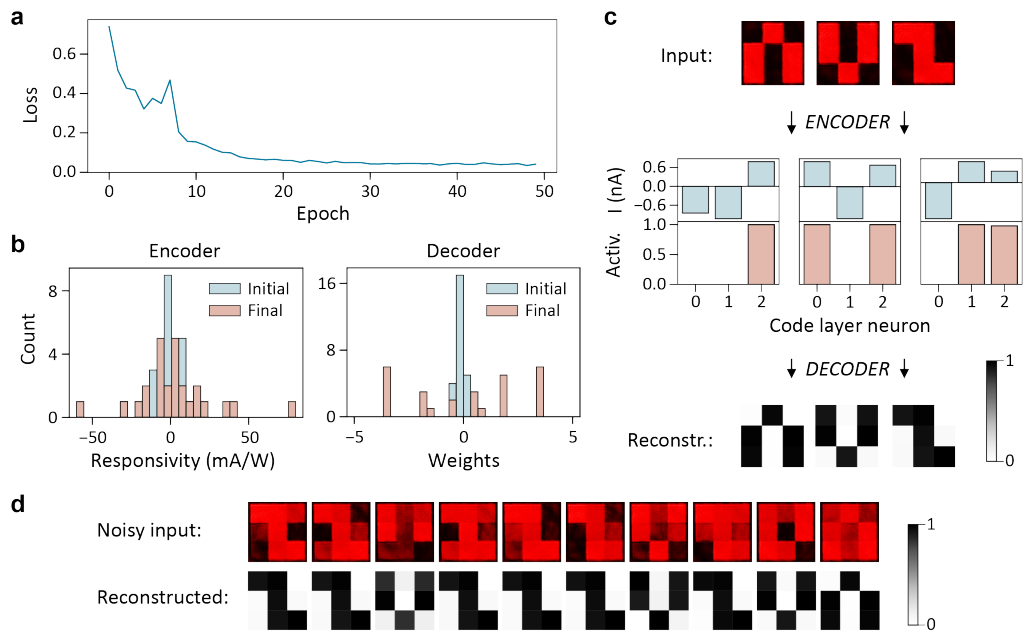

**Figure 4 | Device operation as an autoencoder. a**, Loss of the autoencoder during training. **b**, Responsivity and weight distributions before (initial) and after training (final). **c**, Autoencoding of noise-free letters. The encoder translates the projected images into a current code which is converted by the nonlinearity into a binary activation code and finally reconstructed into an image by the decoder. **d**, Randomly chosen noisy inputs ($\sigma = 0.3$) and the corresponding reconstructions after autoencoding.

As image sensing and processing are both performed in the analog domain, the operation speed of the system is only limited by physical processes involved in the photocurrent generation [28]. As a result, image recognition/encoding occurs in real-time with a rate that is orders of magnitude higher than what can be achieved conventionally. In order to demonstrate the high-speed capabilities of the sensor, we performed measurements with a 40 ns pulsed laser source (522 nm, ~10 W/cm²). The photodiode array was operated as a classifier and trained beforehand, as discussed above. We subsequently projected two letters ('v' and 'n') and measured the time-resolved currents of the two corresponding channels. In Fig. 5 we plot the electric output pulses, demonstrating the correct pattern classification within ~50 ns. The system is thus capable of processing images with a throughput of 20 million bins per second. We emphasize that this value is only limited by the 20 MHz-bandwidth of the used amplifiers and substantially higher rates are possible. Such a network may hence provide new opportunities for ultrafast microscopy. It may also be employed in ultrafast spectroscopy for the detection and classification of spectral events.

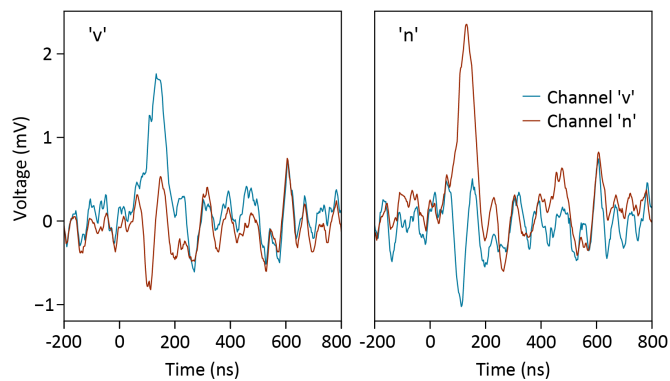

**Figure 5 | Real-time image recognition.** Projection of two different letters 'v' and 'n' with 40 ns duration, leading to distinct output voltages of the labeled channels.

Let us now comment on the prospects for scalability. In our present implementation the weights of the ANN are stored in an external memory and supplied to each detector via cabling. Scaling will require to store the weights locally. This could, for example, be achieved by using ferroelectric gate dielectrics or by employing floating gate devices [29,30,31]. To demonstrate the feasibility of the latter approach, we present in Supplementary Fig. S4 a floating split-gate photodetector. Once set, this detector indeed 'remembers' its responsivity value and delivers a photocurrent of adjustable sign/magnitude. During training, each detector could then be addressed by its column and row using the standard infrastructure of today's active pixel cameras.

Another important question is the number of required subpixels $M$. As shown in the example in Fig. 1d, a segmentation of each pixel into 3×3 subpixels may be adequate for some applications. Given the exponential increase of the network complexity with $M$, increasing the segmentation to 6×6 subpixels would already result in a very powerful ANN with a manageable number of 36 analog outputs. We propose that such a network may also be trained as a binary-hashing [32] autoencoder, eliminating the need for analog-to-digital conversion. In this case, a 36-bit digital output allows for as many as $2^{36} - 1 \approx 7 \times 10^{10}$ encodable features. The implementation of an analog deep learning network becomes feasible by converting the photocurrents into voltages that are then fed into a memristor crossbar. We finally remark that besides on-chip training, demonstrated in this letter, the network can also be trained off-line, using computer simulations, and the predetermined photoresponsivity matrix then loaded into the device.

In conclusion, we have presented an ANN vision sensor for ultrafast recognition and encoding of images. The device concept is easily scalable and provides various training possibilities for time-critical machine vision applications.

## METHODS

**Device fabrication.** The fabrication of the chip followed the procedure described in Ref. [24]. As a substrate we used a silicon wafer with 280 nm-thick $SiO_2$. First, we prepared a bottom metal layer by writing a design with electron-beam lithography (EBL) and evaporating Ti/Au (3/30 nm). Second, we deposited a 30 nm-thick $Al_2O_3$ gate oxide using atomic layer deposition (ALD). Via-holes through the $Al_2O_3$ isolator, which were necessary for connections between top and bottom metal layers, were defined by EBL and etched with 30% solution of KOH in DI water. Third, we mechanically exfoliated a (~70×120 μm² large) $WSe_2$ flake from a bulk crystal (*HQ Graphene*) and transferred it onto the desired position on the sample by an all-dry viscoelastic stamping method [33]. Its thickness (~6 monolayers, or ~4 nm) was estimated from the contrast under which it appears in an optical microscope. Next, we separated 27 pixels from the previously transferred $WSe_2$ sheet by defining a mask with EBL and reactive ion etching (RIE) with Ar/SF6 plasma. Mild RIE oxygen plasma treatment allowed to remove the crust from the surface of the polymer mask that appeared during the preceding etching step. Then, a top metal layer was added by another EBL process and Ti/Au (3/32 nm) evaporation. We confirmed the continuity and solidity of the electrode structure by scanning electron microscopy (SEM) and electrical measurements. The sample was finally mounted in a 68-pin chip carrier and wire-bonded.

**Experimental setup.** A schematic of the experimental setup is shown in Fig. 2d. Light from a semiconductor laser (650 nm wavelength) was linearly polarized before it illuminated a spatial light modulator (SLM; *Hamamatsu*), operated in intensity-modulation mode. On the SLM, the letters were displayed and the polarization of the light was rotated depending on the pixel value. A linear polarizer with its optical axis oriented normal to the polarization of the incident laser light functioned as an analyzer. The so-generated optical image was then projected onto the sample using a 20× microscope objective with long working-distance (*Mitutoyo*). Gate voltage pairs were supplied to each of the detectors individually using a total of 54 digital-to-analog converters (*National Instruments* NI-9264) and the three output currents were measured by source meter units (*Keithley* 2614B). For time-resolved measurements a pulsed laser source, emitting ~40 ns long pulses at 522 nm wavelength, was used. The output current signals were amplified with high-bandwidth (20 MHz) transimpedance amplifiers (*Femto*) and the output voltages were recorded with an oscilloscope (*Keysight*). For the time-resolved measurements, the analyzer was removed and the SLM was operated in phase-only mode to achieve higher illumination intensities (~10 W/cm$^2$). The phase-only Fourier transforms of the projected images were calculated using the Gerchberg-Saxton algorithm [ 34 ]. All measurements were conducted at room temperature and under high-vacuum condition (~$10^{-6}$ mbar).

**Data availability.** The data that support the findings of this study are available from the corresponding authors upon request.

**Acknowledgments:** We thank Benedikt Limbacher for helpful discussions and Andreas Kleinl, Matthias Paur and Fabian Dona for technical assistance. We acknowledge financial support by the Austrian Science Fund FWF (START Y 539-N16) and the European Union (grant agreement No. 785219 Graphene Flagship and Marie Sklodowska-Curie Individual Fellowship OPTOvanderWAALS, grant ID 791536).

**Author contributions:** T.M. conceived the experiment. L.M. designed and built the experimental setup, programmed the machine learning algorithms, carried out the measurements and analyzed the data. J.S. fabricated the ANN vision sensor. S.W. and D.K.P. contributed to the sample fabrication. A.J.M.-M. fabricated and characterized the floating-gate detector. L.M., J.S. and T.M. prepared the manuscript. All authors discussed the results and commented on the manuscript.

**Competing financial interests:** The authors declare no competing financial interests.